\documentclass{article}
\usepackage{spconf,amsmath,graphicx,hyperref}

\usepackage{wrapfig}
\usepackage{graphicx}
\usepackage{subfigure}
\usepackage{multirow, makecell}
\usepackage{enumitem}
\usepackage{booktabs}
\usepackage{amsfonts}
\usepackage{cite}
\usepackage[capitalize,noabbrev]{cleveref}

\title{ZipCodec: Ultra-Low-Frame-Rate Streaming Speech Coding}
\name{Luca Della Libera$^{1,2}$, Cem Subakan$^{3,1,2}$, Mirco Ravanelli$^{1,2}$
}
\address{\textit{$^1$Concordia University, 
  $^2$Mila-Quebec AI Institute,
  $^3$Université Laval}
}

\begin{document}
\ninept
\maketitle
\begin{abstract}
Neural audio codecs are a fundamental component of modern speech generation systems. While recent codecs achieve increasingly low bitrates, reducing frame rate remains challenging, as each token must preserve more information while maintaining reconstruction quality. We present ZipCodec, a streaming neural speech codec operating at 6.25 Hz and 0.80 kbps with a theoretical latency of 160 ms. Our approach combines large-scale WavLM distillation with a redesigned transformer-based architecture, a scalar spherical quantizer, and a latency-aware streaming decoder. Experiments show that ZipCodec substantially outperforms existing streaming codecs at comparable bitrates in both reconstruction and downstream tasks, while operating at a significantly lower frame rate. Despite its 842M parameters, ZipCodec achieves real-time single-stream inference on a consumer-grade CPU. {Demo samples, code and checkpoints are available at \href{https://lucadellalib.github.io/zipcodec-web/}{https://lucadellalib.github.io/zipcodec-web/}.}
\end{abstract}

\section{Introduction}
\label{sec:intro}
Neural audio codecs~\cite{zeghidour2021soundstream, defossez2023encodec, kumar2023dac} have become a key component of modern speech generation systems, providing compact discrete representations of speech that can be modeled autoregressively. Building on the success of large language models~\cite{dubey2024llama3herdmodels,jiang2024mixtralexperts,comanici2025gemini25,singh2025openaigpt5card,deepseekai2025deepseekv3}, this discrete-token paradigm has been extended from text to speech, enabling a new generation of speech-native language models~\cite{hassid2023textually,defossez2024moshi,nguyen2024spiritlminterleavedspokenwritten,park2024speechssm,dellalibera2026wavslm}.

Recent neural codec research has explored several directions towards more compact and expressive speech representations, including lower frame rates, single-codebook designs, semantic distillation, and supervised fine-tuning~\cite{zhang2024speechtokenizer,parker2024scaling,casanova2025nano,ji2024wavtokenizer,xin2024bigcodec,wu2024ts3codec,dellalibera2025focalcodec,ye2025llasa,dellalibera2025focalcodecstream,paissan2026exploring}. However, simultaneously achieving a low bitrate, rich semantic and acoustic representations, high reconstruction quality, and streamability remains challenging.

Among the factors determining bitrate, frame rate is arguably the most critical for speech language modeling, as it directly determines the length of the resulting token sequence. Reducing the frame rate therefore shortens the sequence, lowering computational cost and simplifying sequence modeling. At the same time, it creates an increasingly severe information bottleneck: each discrete token must encode linguistic content, speaker characteristics, prosody, and fine acoustic details over a longer temporal interval.

Recent codecs have shown that extremely low frame rates are possible when relaxing some of these requirements. U-Codec~\cite{yang2025ucodec}, for example, operates at 5 Hz but is designed for offline acoustic reconstruction. TaDiCodec~\cite{wang2025tadicodec} reaches 6.25 Hz by leveraging text as additional side information for speech reconstruction. FlexiCodec~\cite{li2025flexicodec} and DyCAST~\cite{dellalibera2026dycast} instead adopt variable-frame-rate representations that can reach similarly low average frame rates, but operate offline and exhibit increasing reconstruction degradation as the frame rate is reduced. Among streaming codecs that jointly capture semantic and acoustic information, Mimi~\cite{defossez2024moshi} operates at 12.5 Hz. To the best of our knowledge, no such codec has been demonstrated below 12.5 Hz, leaving open how far the frame rate of a streaming codec can be reduced.

In this work, we show that this limit can be pushed further by introducing \textbf{ZipCodec}, a streaming speech codec operating at 6.25 Hz. Building on FocalCodec-Stream~\cite{dellalibera2025focalcodecstream}, ZipCodec distills WavLM~\cite{chen2022wavlm} layer-6 representations into a causal backbone, while rethinking the codec architecture and further scaling both model capacity and training data. These advances enable ZipCodec to compress speech at 0.80 kbps while substantially improving both reconstruction and representation quality. Despite its extremely low frame rate, ZipCodec remains fully streamable, with a theoretical latency of 160 ms, which is below the 200 ms timescale of typical conversational turn transitions~\cite{levinson2015timing}, making the proposed representations compatible with highly responsive streaming speech-to-speech systems.
Our main contributions are as follows:

\begin{itemize}[topsep=0.2cm, itemsep=0.2cm, leftmargin=0.35cm]
    \item We introduce ZipCodec, a \textbf{streaming} neural speech codec that compresses speech at an exceptionally low frame rate of \textbf{6.25 Hz} and a bitrate of \textbf{0.80 kbps}, with a theoretical latency of 160~ms. Despite its \textbf{842M parameters}, ZipCodec supports real-time inference for a single stream on a consumer-grade CPU.
    \item To achieve this, we substantially redesign the FocalCodec-Stream architecture for improved scalability and efficiency. In particular, we adopt an optimized transformer-based architecture that removes normalization layers and positional encodings, employ scalar spherical quantization~\cite{dellalibera2026dycast} to obtain a compact factorized bottleneck, and introduce a latency-aware streaming decoder. We further scale WavLM layer-6 distillation to approximately \textbf{94,000 hours of English speech} and reproduce WavLM noise and overlapping-speech augmentation strategy to better match WavLM original data distribution.
    \item We extensively evaluate ZipCodec across reconstruction and downstream tasks, demonstrating substantial improvements in both reconstruction and representation quality over other streaming codecs at matched bitrates, despite operating at a significantly lower frame rate.
\end{itemize}

\section{ZipCodec}
\label{sec:method}

\begin{figure}[t!]
  \centering
\includegraphics[width=0.49\textwidth]{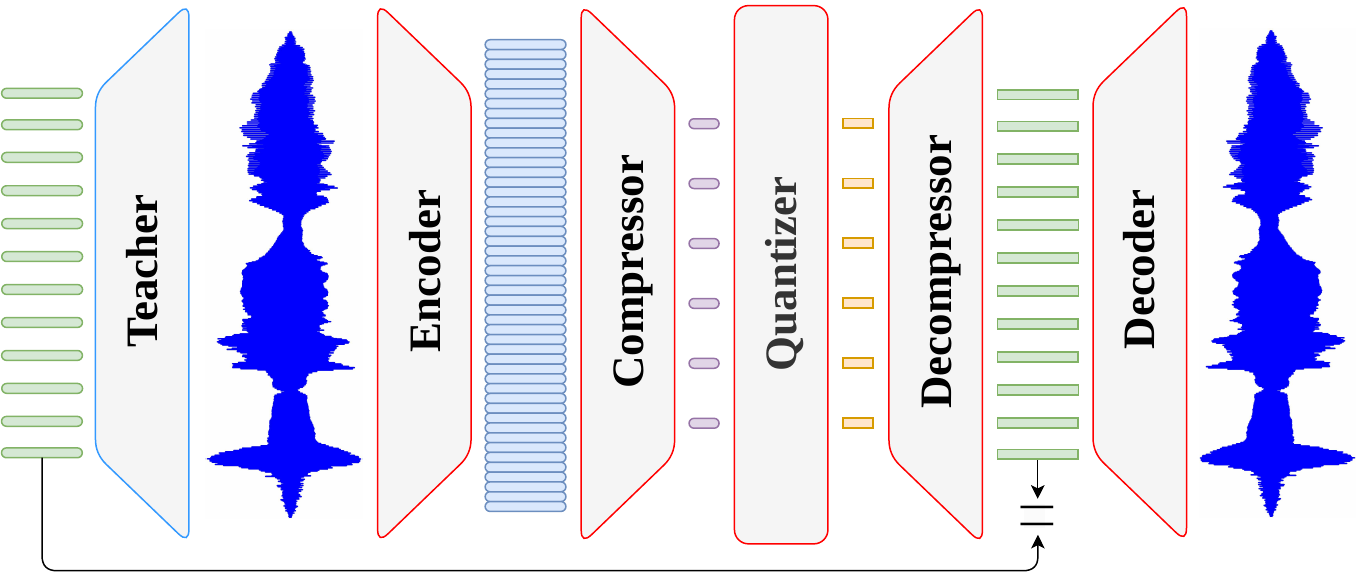}
  \vspace{-0.6cm}
  \caption{ZipCodec architecture. The encoder extracts features containing both acoustic and semantic information. These features are then mapped to a low-dimensional space by the compressor, quantized, and projected back by the decompressor. The decoder resynthesizes the waveform from these features. All these modules are causal, while a non-causal teacher is used for distillation to align causal features with their non-causal counterparts.}
  \label{fig:zipcodec}
\vskip -0.4cm
\end{figure}

\subsection{Architecture}
\label{sec:architecture}
Our streaming codec builds upon the FocalCodec-Stream~\cite{dellalibera2025focalcodecstream} architecture, with several modifications designed to improve scalability and reconstruction quality. Following the same overall structure, ZipCodec consists of five main components (see \cref{fig:zipcodec}): \emph{encoder}, \emph{compressor}, \emph{quantizer}, \emph{decompressor}, and \emph{decoder}.

\begin{description}[leftmargin=0.0cm]
\item[Encoder.] In contrast to FocalCodec-Stream, which employs a learned convolutional encoder followed by transformer blocks, ZipCodec uses a causal log-mel frontend. We extract 80-dimensional log-mel features using a 25 ms Hann window and a 10 ms hop, resulting in a 100 Hz feature sequence. This simple frontend removes the need for a learned waveform encoder while providing compact representations to the subsequent compressor.

\item[Compressor.]
The compressor first employs a \emph{temporal patching} module that groups 16 consecutive log-mel frames and linearly projects them to the model dimension. This reduces the frame rate by a factor of 16, from 100 Hz to 6.25 Hz, with each representation spanning 160 ms of speech. The resulting sequence is then processed by a causal transformer-based backbone.

While FocalCodec-Stream relies on focal modulation~\cite{yang2022focalnets,dellalibera2024focal} as its main sequence modeling operator, we adopt a transformer-based design to leverage the highly optimized attention and matrix-multiplication primitives available on modern hardware.
Rather than using a standard transformer, we introduce \emph{ErfFormer}, which builds on LLaMA-style blocks~\cite{dubey2024llama3herdmodels} with grouped-query attention and gated feed-forward networks with SiLU activations, while making two key modifications for efficient streaming speech modeling: eliminating both normalization layers and positional encodings.
Specifically, ErfFormer replaces RMSNorm~\cite{zhang2019root} with DynamicErf~\cite{chen2025stronger}, a lightweight activation that has been shown to match or outperform normalization-based architectures at lower computational cost. ErfFormer also removes positional encodings entirely. Unlike discrete text tokens, continuous acoustic representations already carry local temporal structure, while causal attention preserves their temporal ordering. More importantly, removing position-dependent representations facilitates long-running streaming inference, allowing the model to operate beyond the context lengths observed during training without requiring positional extrapolation.

\item[Quantizer.]
We discretize the compressor output using scalar spherical quantization (SSQ)~\cite{dellalibera2026dycast}. Given a compressor representation, we first project it to an $L=64$-dimensional latent space and normalize it to the unit hypersphere. Each latent dimension is then independently quantized to one of $K=4$ uniformly spaced scalar levels in the interval $[-1/\sqrt{L}, \, \, 1/\sqrt{L}]$. The resulting representation contains 64 two-bit symbols per frame, corresponding to a bitrate of 0.80 kbps at a frame rate of 6.25 Hz. Despite its factorized structure, SSQ implicitly defines $K^L=4^{64}$ possible joint codewords without requiring an explicit codebook of this size. The quantized representation is then renormalized to the unit hypersphere and projected back to the model dimension before being passed to the decompressor.

\item[Decompressor.]
The decompressor mirrors the architecture of the compressor, employing the same ErfFormer backbone followed by a \emph{temporal unpatching} module. The quantized representations are first processed by ErfFormer and then linearly projected and unpatched, expanding each 6.25 Hz representation into 8 1024-dimensional WavLM layer-6 representations at 50 Hz. Thus, each 160 ms discrete representation is mapped to a sequence of 8 continuous representations spaced 20 ms apart, which are then passed to the waveform decoder.

\item[Decoder.]
The 50 Hz WavLM representations are converted to waveform samples using a streaming Vocos~\cite{siuzdak2023vocos} decoder. We adapt its inverse STFT synthesis to streaming inference by maintaining the overlap-add state across consecutive decoding steps.
The exceptionally low frame rate of ZipCodec also allows us to relax the causality constraints within each decoding step. Since all 8 corresponding WavLM representations are available within the same 160 ms interval, we process them jointly rather than restricting the decoder to causal processing at the 20 ms feature granularity. Specifically, we use \emph{left and right} convolutional padding while constraining the overall receptive field to the 160 ms decoding window. This allows the decoder to leverage future context within each step without increasing the theoretical latency imposed by the 6.25 Hz bottleneck.
\end{description}

\subsection{Training}
\label{sec:training}
ZipCodec follows a similar training strategy to FocalCodec-Stream, in which the encoder-compressor-quantizer-decompressor pipeline is trained by distilling WavLM features, while the waveform decoder is trained separately on continuous WavLM representations. However, we considerably simplify the distillation procedure: rather than the four-stage approach used in FocalCodec-Stream, we jointly train the encoder, compressor, quantizer, and decompressor in a single stage to reconstruct continuous WavLM layer-6 representations. We further scale both model capacity and training data, using approximately 94,000 hours of speech from LibriLight~\cite{kahn2020libri}, VoxPopuli~\cite{wang2021voxpopuli}, and GigaSpeech~\cite{chen2021gigaspeech}, closely matching the data distribution used for WavLM pretraining.

To further reduce the distribution gap between WavLM pretraining and distillation, we reproduce its noise and overlapping-speech augmentation strategy. Each training utterance is augmented with probability 0.2 by mixing a randomly selected segment with either another utterance, simulating overlapping speech, or noise from the DNS~\cite{reddy21_interspeech} dataset. Conditioned on augmentation, DNS noise is selected with probability 0.1 and mixed at an energy ratio uniformly sampled between $-5$ and $20$~dB; otherwise, another training utterance is mixed at a ratio between $-5$ and $5$~dB. Importantly, the same augmented waveform is provided to both ZipCodec and the frozen WavLM teacher, such that ZipCodec learns to reconstruct the WavLM representations of the augmented speech itself.

\section{Experimental Setup}
\label{sec:setup}
We train the ZipCodec encoder, compressor, quantizer, and decompressor on 4 NVIDIA H100 (80 GB) GPUs with a batch size of 4 per GPU, for a total batch size of 16. Training examples consist of 40.96 s segments sampled from LibriLight, VoxPopuli, and GigaSpeech proportionally to their approximate corpus sizes. Utterances longer than 40.96 s are randomly cropped, while shorter utterances are repeated and randomly cropped to the target duration; utterances shorter than 2 s are discarded. Training shards are continuously resampled and shuffled rather than traversed in fixed epochs. In addition to the main L2 reconstruction loss, we employ an entropy loss to encourage high utilization of the scalar quantization levels. We use AdamW~\cite{loshchilov2019adamw} with $\beta_1=0.9$, $\beta_2=0.98$, a peak learning rate of $2\times10^{-4}$, and a weight decay of $0.01$. The learning rate is linearly warmed up for 10,000 optimization steps and then decayed to $2\times10^{-5}$ following a cosine schedule. Gradients are clipped to a global norm of $1.0$, and training is performed using bfloat16 mixed precision for a total of 4M optimization steps.

The compressor consists of 6 ErfFormer blocks with a model dimension of 2048 and a feed-forward dimension of 8192. Attention employs 16 query heads and 4 key-value heads, each with a head dimension of 128. The decompressor mirrors this architecture, using the same number of layers and dimensions. During streaming inference, both modules maintain a bounded key-value cache of 256 frames, corresponding to 40.96 s of context at 6.25 Hz and matching the context length used during training. For arbitrarily long streams, we reset the cache every 256 frames rather than using a sliding window, which would expose the model to attention patterns across longer sequences not encountered during training.

For the decoder, we build upon the original Vocos training recipe and train on LibriTTS-100~\cite{zen2019libritts}, resampled to 16 kHz. The decoder consists of 20 ConvNeXt blocks with a hidden dimension of 1024 and a kernel size of 7, followed by a complex spectral head and inverse STFT synthesis. All temporal components maintain explicit streaming state, including the convolutional and overlap-add states. At each streaming step, the decoder consumes 8 WavLM representations and produces 2,560 waveform samples, corresponding to 160 ms of audio. We employ the multi-scale and multi-period discriminators from \cite{kong2020hifigan}, together with the multi-resolution discriminator from \cite{kumar2023dac}. We additionally incorporate the speaker consistency loss proposed in \cite{casanova2025nano}, using WavLM-base-SV~\cite{chen2022wavlm} as a pretrained speaker embedding extractor. We train on 7,040-sample audio segments with a batch size of 16 using AdamW with $\beta_1=0.8$, $\beta_2=0.99$, an initial learning rate of $2\times10^{-4}$, and a weight decay of $0.01$. The learning rate follows an exponential decay schedule with a factor of $0.999$. Training continues until perceptual quality saturates, which occurs after approximately 5M optimization steps.

\begin{table}[t!]
\setlength{\tabcolsep}{3pt}
\vspace{-6pt}
\caption{Codecs considered in our experiments.}
\label{tab:baselines}
\centering
\resizebox{0.49\textwidth}{!}{%
\begin{tabular}{lcccccc}
\toprule
\textbf{Codec} &
\makecell{\textbf{Frame Rate} \\ \textbf{(Hz)}} &
\makecell{\textbf{Bitrate} \\ \textbf{(kbps)}} &
\makecell{\textbf{Sample Rate} \\ \textbf{(kHz)}} &
\textbf{Codebooks} &
\makecell{\textbf{Latency} \\ \textbf{(ms)}} &
\makecell{\textbf{Params} \\ \textbf{(M)}} \\
\midrule
EnCodec          & 75.0 & 1.50 & 24      & 2 $\times$ 1024  & 13  & 15  \\
AudioDec         & 80.0 & 1.60 & 24      & 2 $\times$ 1024  & 13  & 8   \\
HILCodec         & 75.0 & 1.50 & 24      & 2 $\times$ 1024  & 13  & 11  \\
Mimi             & 12.5 & 0.83 & 24      & 6 $\times$ 2048  & 80  & 82  \\
PAST             & 50.0 & 1.00 & 16      & 2 $\times$ 1024  & 20  & 126 \\
FocalCodec-S@50  & 50.0 & 0.80 & 16 / 24 & 1 $\times$ 65536 & 80  & 249 \\
\midrule
\textbf{ZipCodec}         & 6.25 & 0.80 & 16      & 64 $\times$ 4    & 160 & 842 \\
\midrule
FocalCodec@50    & 50.0 & 0.65 & 16      & 1 $\times$ 8192  & --- & 142 \\
\bottomrule
\end{tabular}
}
\vskip -0.2in
\end{table}

\section{Results}
\label{sec:results}
We adopt the evaluation protocol of FocalCodec-Stream~\cite{dellalibera2025focalcodecstream}, focusing on streaming codecs operating in the low-bitrate regime. For models supporting multiple quantizer configurations, we select the setting closest to ZipCodec bitrate of 0.80 kbps to enable comparisons under similar compression constraints. Our baselines include the acoustic codecs EnCodec~\cite{defossez2023encodec}, AudioDec~\cite{wu2023audiodec}, and HILCodec~\cite{ahn2024hilcodec}, as well as the hybrid codecs Mimi~\cite{defossez2024moshi} and PAST~\cite{hartuv2025past}, which incorporate semantic information through distillation and supervised fine-tuning, respectively. We further compare against FocalCodec-Stream, the closest baseline to ZipCodec in terms of system design. Finally, we include the original non-streaming FocalCodec@50~\cite{dellalibera2025focalcodec} as an offline reference for WavLM layer-6 distillation. The configurations of all evaluated codecs are summarized in \cref{tab:baselines}.

\subsection{Speech Resynthesis and Voice Conversion}
We first evaluate ZipCodec on speech resynthesis (\textbf{SR}) in both {English} and {multilingual} settings, using the evaluation protocol introduced in \cite{dellalibera2025focalcodec}. We evaluate English resynthesis on LibriSpeech~\cite{panayotov2015librispeech} \texttt{test-clean} and multilingual resynthesis on a subset of MLS~\cite{pratap2020mls}. Reconstruction quality is measured along several dimensions. {UTMOS}~\cite{saeki2022utmos} evaluates perceptual naturalness, while intelligibility is assessed through {dWER}, computed as the WER between Whisper-small~\cite{radford2022robust} transcriptions of the original and reconstructed speech. Speaker preservation is quantified using WavLM-based embedding similarity ({Sim}). We additionally report {code usage} and {normalized entropy} to quantify codebook utilization, as well as the real-time factor (RTF) to measure inference efficiency. RTF is measured on a 1/8 partition of an NVIDIA H100 (80 GB) GPU using multi-instance GPU partitioning.

As shown in \cref{tab:speech_resynthesis}, ZipCodec achieves the strongest overall speech resynthesis performance among streaming codecs in both the English and multilingual settings, consistently improving over FocalCodec-Stream in perceptual quality, intelligibility, and speaker fidelity. The improvements are particularly pronounced in the multilingual setting, despite ZipCodec operating at only 6.25 Hz compared to 50 Hz for FocalCodec-Stream and at least 12.5 Hz for the other streaming baselines. ZipCodec also achieves full code utilization in both settings while maintaining high normalized entropy, indicating effective use of the factorized SSQ bottleneck. Finally, despite its substantially lower frame rate, ZipCodec approaches the reconstruction quality of the non-streaming FocalCodec@50 reference, narrowing the gap in perceptual quality and intelligibility while surpassing it in speaker fidelity.

We also perform one-shot voice conversion (\textbf{VC}) experiments to assess the ability of ZipCodec to disentangle linguistic content from speaker information. Following the protocol of \cite{dellalibera2025focalcodecstream}, we use a dataset of parallel utterances derived from VCTK~\cite{veaux2017cstr}.
As shown in \cref{tab:speech_resynthesis}, ZipCodec achieves the highest perceptual quality among streaming codecs while maintaining strong speaker fidelity, second only to FocalCodec-Stream. It also preserves competitive intelligibility, outperforming most streaming baselines, with only FocalCodec-Stream and PAST achieving lower dWER. Overall, these results show that the 6.25 Hz bottleneck retains sufficient information for effective voice conversion despite its aggressive temporal compression.

\begin{table*}[t!]
\setlength{\tabcolsep}{2pt} 
\vspace{-6pt}
\caption{Speech resynthesis and voice conversion. \textbf{Best}, \underline{second-best} and \setlength{\fboxsep}{1pt}\fbox{{best non-streaming}} results are highlighted.}
\label{tab:speech_resynthesis}
\begin{center}
\begin{footnotesize}
\resizebox{0.99\textwidth}{!}{%
\begin{tabular}{lcc|cccccc|cccccc|ccc}
\toprule
\multirow{4}{*}{\textbf{Codec}} &
\multirow{4}{*}{\makecell{\textbf{Frame} \\ \textbf{Rate (Hz)}}} &
\multirow{4}{*}{\makecell{\textbf{Bitrate} \\ \textbf{(kbps)}} \,}
& \multicolumn{6}{c|}{\textbf{SR -- English}}
& \multicolumn{6}{c|}{\textbf{SR -- Multilingual}}
& \multicolumn{3}{c}{\textbf{VC}} \\
\cmidrule{4-18}
& & 
& \textbf{UTMOS} $\uparrow$ & \textbf{dWER} $\downarrow$ & \textbf{Sim} $\uparrow$ & \makecell{\textbf{Code} \\ \textbf{Usage}} $\uparrow$ & \makecell{\textbf{Norm.} \\ \textbf{Entropy}} $\uparrow$ & \textbf{RTF} $\uparrow$
& \textbf{UTMOS} $\uparrow$ & \textbf{dWER} $\downarrow$ & \textbf{Sim} $\uparrow$ & \makecell{\textbf{Code} \\ \textbf{Usage}} $\uparrow$ & \makecell{\textbf{Norm.} \\ \textbf{Entropy}} $\uparrow$ & \textbf{RTF} $\uparrow$
& \textbf{UTMOS} $\uparrow$ & \textbf{dWER} $\downarrow$ & \textbf{Sim} $\uparrow$ \\
\midrule
\multicolumn{1}{l}{Reference} & --- & ---
& 4.09 & 0.00 & 100.0 & --- & --- & ---
& 2.84 & 0.00 & 100.0 & --- & --- & ---
& 4.09 & 0.00 & 100.0 \\

\multicolumn{1}{l}{EnCodec} & 75.0 & 1.50
& 1.58 & 8.08 & 93.8 & 93.4 & 82.1 & 91
& 1.33 & 29.60 & 95.5 & 93.4 & 79.2 & 113
& 1.24 & 86.52 & 72.2 \\

\multicolumn{1}{l}{AudioDec} & 80.0 & 1.60
& 1.48 & 11.61 & 92.1 & 91.9 & 70.0 & \underline{145}
& 1.29 & 40.95 & 92.3 & 87.5 & 68.2 & \underline{195}
& 1.26 & 68.45 & 68.2 \\

\multicolumn{1}{l}{HILCodec} & 75.0 & 1.50
& 2.86 & 6.65 & 95.4 & \underline{99.0} & 95.6 & 41
& 1.81 & 25.32 & 97.8 & 99.1 & 94.8 & 41
& 1.40 & 58.36 & 76.8 \\

\multicolumn{1}{l}{Mimi} & 12.5 & 0.83
& 3.44 & 4.77 & 96.6 & 96.2 & 92.0 & \textbf{154}
& 2.19 & 26.12 & 97.4 & 96.5 & 89.2 & \textbf{216}
& 2.62 & 110.00 & 91.3 \\

\multicolumn{1}{l}{PAST} & 50.0 & 1.00
& 2.33 & 4.04 & 83.8 & 56.7 & 90.7 & 59
& 1.44 & 49.35 & 80.8 & 57.0 & 87.5 & 63
& 1.42 & \textbf{18.28} & 68.5 \\

\multicolumn{1}{l}{FocalCodec-S@50} & 50.0 & 0.80
& \underline{3.85} & \underline{3.68} & \underline{97.0} & \textbf{100.0} & \textbf{98.7} & 106
& \underline{2.65} & \underline{19.88} & \underline{98.1} & \underline{99.2} & \textbf{98.3} & 107
& \underline{3.10} & \underline{22.71} & \textbf{92.5} \\
\midrule

\multicolumn{1}{l}{\textbf{ZipCodec}} & 6.25 & 0.80
& \textbf{3.89} & \textbf{2.83} & \textbf{98.0} & \textbf{100.0} & \underline{96.2} & 62
& \textbf{2.69} & \textbf{15.52} & \textbf{98.7} & \textbf{100.0} & \underline{96.5} & 106
& \textbf{3.15} & 25.90 & \underline{91.5} \\
\midrule

\multicolumn{1}{l}{FocalCodec@50} & 50.0 & 0.65
& \setlength{\fboxsep}{1pt}\fbox{{4.05}} & \setlength{\fboxsep}{1pt}\fbox{{2.18}} & 97.4 & \setlength{\fboxsep}{1pt}\fbox{{100.0}} & \setlength{\fboxsep}{1pt}\fbox{{98.9}} & 123
& \setlength{\fboxsep}{1pt}\fbox{{2.96}} & \setlength{\fboxsep}{1pt}\fbox{{12.57}} & 98.3 & \setlength{\fboxsep}{1pt}\fbox{{100.0}} & 98.1 & 116
& \setlength{\fboxsep}{1pt}\fbox{{3.38}} & 21.27 & 92.2 \\
\bottomrule
\end{tabular}
}
\end{footnotesize}
\end{center}
\vskip -0.3in
\end{table*}

\begin{table*}[t!]
\setlength{\tabcolsep}{6pt}
\vspace{-3pt}
\caption{Discriminative and generative downstream tasks. \textbf{Best}, \underline{second-best} and \setlength{\fboxsep}{1pt}\fbox{{best non-streaming}} results are highlighted.}
\label{tab:downstream}
\begin{center}
\begin{scriptsize}
\resizebox{0.99\textwidth}{!}{%
\begin{tabular}{lcc|c|c|c|c|c|ccc|ccc}
\toprule
\multirow{3}{*}{\textbf{Codec}} &
\multirow{3}{*}{\makecell{\textbf{Frame} \\ \textbf{Rate (Hz)}}} &
\multirow{3}{*}{\makecell{\textbf{Bitrate} \\ \textbf{(kbps)}} \,}
& \multicolumn{1}{c|}{\textbf{ASR}}
& \multicolumn{1}{c|}{\textbf{SI}}
& \multicolumn{1}{c|}{\textbf{SER}}
& \multicolumn{1}{c|}{\textbf{KS}}
& \multicolumn{1}{c|}{\textbf{IC}}
& \multicolumn{3}{c|}{\textbf{SE}}
& \multicolumn{3}{c}{\textbf{SS}} \\
\cmidrule{4-14}
& & &
\textbf{WER} $\downarrow$ & \textbf{ER} $\downarrow$ & \textbf{ER} $\downarrow$ & \textbf{ER} $\downarrow$ & \textbf{ER} $\downarrow$
& \textbf{DNSMOS} $\uparrow$ & \textbf{dWER} $\downarrow$ & \makecell{\textbf{Sim}} $\uparrow$
& \textbf{DNSMOS} $\uparrow$ & \textbf{dWER} $\downarrow$ & \makecell{\textbf{Sim}} $\uparrow$ \\
\midrule
\multicolumn{1}{l}{Reference} & --- & ---
& --- & --- & --- & --- & ---
& 3.56 & 0.00 & 100.0 & 3.77 & 0.00 & 100.0 \\

\multicolumn{1}{l}{EnCodec} & 75.0 & 1.50
& 28.55 & 3.25 & 41.94 & 96.16 & 49.79
& 3.13 & 37.31 & 85.6 & 3.11 & {77.61} & 87.4 \\

\multicolumn{1}{l}{AudioDec} & 80.0 & 1.60
& 29.21 & \underline{1.69} & 45.85 & 25.30 & 46.77
& 2.96 & 61.11 & 84.3 & 2.97 & {88.59} & 84.0 \\

\multicolumn{1}{l}{HILCodec} & 75.0 & 1.50
& 29.89 & {1.98} & 51.61 & {15.17} & 53.69
& 3.32 & 41.33 & \underline{90.2} & 3.35 & 78.43 & 86.9 \\

\multicolumn{1}{l}{Mimi} & 12.5 & 0.83
& 22.56 & {3.13} & {35.71} & {5.81} & 35.74
& 3.14 & 55.99 & 86.7 & 3.32 & 86.49 & 88.9 \\

\multicolumn{1}{l}{PAST} & 50.0 & 1.00
& \textbf{10.74} & 3.43 & {36.41} & 6.41 & {31.66}
& 3.15 & \textbf{18.19} & 77.9 & 3.15 & 85.61 & 80.3 \\

\multicolumn{1}{l}{{FocalCodec-S@50}} & 50.0 & 0.80
& {17.02} & {2.18} & \underline{34.56} & \underline{5.63} & \underline{29.49}
& \underline{3.56} & \underline{19.56} & 87.7 & \underline{3.68} & \underline{75.43} & \underline{90.8} \\
\midrule

\multicolumn{1}{l}{\textbf{ZipCodec}} & 6.25{$^\dagger$} & 0.80
& \underline{16.07} & \textbf{0.49} & \textbf{33.64} & \textbf{3.95} & \textbf{26.76}
& \textbf{3.60} & 20.10 & \textbf{91.0} & \textbf{3.77} & \textbf{66.67} & \textbf{91.0} \\
\midrule

\multicolumn{1}{l}{FocalCodec@50} & 50.0 & 0.65
& 15.33 & \setlength{\fboxsep}{1pt}\fbox{{0.35}} & 34.79 & 4.23 & \setlength{\fboxsep}{1pt}\fbox{{24.66}}
& 3.52 & \setlength{\fboxsep}{1pt}\fbox{{12.35}} & 90.4 & 3.71 & 72.61 & 89.5 \\
\bottomrule
\end{tabular}
}
\end{scriptsize}
\begin{minipage}{0.99\textwidth}
\scriptsize
$^\dagger$Discriminative tasks use the 50 Hz representations after temporal unpatching, while generative tasks use the 6.25 Hz representations before temporal unpatching.
\end{minipage}
\end{center}
\vspace{-0.3in}
\end{table*}

\begin{table}[t!]
\setlength{\tabcolsep}{6pt}
\vspace{-10pt}
\caption{Streaming efficiency of ZipCodec.}
\vspace{2pt}
\label{tab:streaming_efficiency}
\centering
\resizebox{0.48\textwidth}{!}{%
\begin{tabular}{ccccc}
\toprule
\textbf{Device} &
\textbf{Batch Size} &
\textbf{RTF} $\uparrow$ &
\makecell{\textbf{Latency p99}  \textbf{(ms)} $\downarrow$} &
\makecell{\textbf{VRAM}  \textbf{(GiB)} $\downarrow$} \\
\midrule
CPU & 1  & 1.33  & 124.68 & --- \\
\midrule
GPU & 1  & 13.96 & 12.71 & 3.33 \\
GPU & 2  & 11.43 & 15.44 & 3.35 \\
GPU & 4  & 9.29  & 18.86 & 3.40 \\
GPU & 8  & 5.85  & 28.04 & 3.56 \\
GPU & 16 & 4.49  & 43.41 & 3.84 \\
\bottomrule
\end{tabular}
}
\vskip -0.2in
\end{table}

\subsection{Downstream Tasks}
To evaluate the quality of the learned representations beyond reconstruction, we consider the same downstream tasks and experimental protocol as FocalCodec-Stream~\cite{dellalibera2025focalcodecstream}, following the DASB benchmark~\cite{mousavi2024dasb}. We evaluate five discriminative tasks: automatic speech recognition (\textbf{ASR}) and speaker identification (\textbf{SI}) on LibriSpeech-460, speech emotion recognition (\textbf{SER}) on IEMOCAP~\cite{busso2008iemocap}, keyword spotting (\textbf{KS}) on Speech Commands~\cite{warden2018speechcommands}, and intent classification (\textbf{IC}) on SLURP~\cite{bastianelli2020slurp}. We additionally consider two generative tasks: speech enhancement (\textbf{SE}) on VoiceBank~\cite{valentinibotinhao2016voicebank} and speech separation (\textbf{SS}) on Libri2Mix-100~\cite{cosentino2020librimix}. We use the same shallow LSTM-based probes for discriminative tasks and non-autoregressive Conformer models for the generative tasks.
For discriminative tasks, we use the representations reconstructed \emph{after the quantization bottleneck and before the decoder}. In ZipCodec, these correspond to the 50 Hz representations produced by the decompressor. For generative tasks, we instead use the ZipCodec representations before temporal unpatching, allowing the SE and SS models to operate directly at the native 6.25 Hz frame rate and benefit from the short sequence length provided by its low-rate bottleneck.
We report error rates for the discriminative tasks and DNSMOS~\cite{reddy2022dnsmos}, dWER, and speaker similarity for SE and SS. We refer to \cite{dellalibera2025focalcodecstream, dellalibera2025focalcodec} for further details on the downstream evaluation setup.

Results are reported in \cref{tab:downstream}. On discriminative tasks, ZipCodec achieves the best performance among streaming codecs on SI, SER, KS, and IC, while obtaining the second-best ASR result. It consistently improves over FocalCodec-Stream across all five tasks, with particularly large gains in SI and KS. ZipCodec also matches or surpasses the non-streaming FocalCodec@50 reference on SER and KS, while remaining competitive on the other tasks. These results show that the representations reconstructed from the 6.25 Hz bottleneck retain rich linguistic, speaker, and paralinguistic information.

The generative results further demonstrate the effectiveness of the low-frame-rate representation. ZipCodec achieves the strongest overall performance among streaming codecs on both tasks. For SE, it achieves the highest perceptual quality and speaker similarity while maintaining competitive intelligibility. For SS, it consistently outperforms both FocalCodec-Stream and the non-streaming baseline. These results are particularly promising for downstream generative modeling, as ZipCodec combines strong representation quality with sequences that are 8 times shorter than the 50 Hz representations used by FocalCodec-Stream.


\subsection{Streaming Efficiency}
We evaluate the streaming efficiency of ZipCodec on a machine equipped with an Intel i7-10875H CPU with 8 cores @ 2.30 GHz, 32 GB of RAM, and an NVIDIA GeForce RTX 3070 (8 GB) GPU. Measurements are performed on 40.96 s sequences, corresponding to the maximum context maintained by ZipCodec before resetting the key-value cache. We report RTF, together with the 99th-percentile latency of each 160 ms streaming step and peak GPU memory consumption. We evaluate batch size 1 on both CPU and GPU and additionally vary the GPU batch size to assess efficiency under concurrent streams. The reported metrics exclude system-level overheads such as audio capture and playback buffering, host-device transfers, resampling, and network transport.

As shown in \cref{tab:streaming_efficiency}, despite its 842M parameters, ZipCodec supports real-time single-stream inference on a consumer-grade CPU, achieving an RTF of 1.33 with a p99 latency below the 160 ms codec frame duration. GPU inference is substantially faster and scales efficiently to concurrent streams, maintaining real-time performance even at a batch size of 16 while requiring only a modest increase in memory. These results show that the low frame rate of ZipCodec enables efficient streaming despite its large model size.

\section{Conclusion}
\label{sec:conclusions}
We introduced ZipCodec, a streaming neural speech codec operating at 6.25 Hz and 0.80 kbps with a theoretical latency of 160 ms. By combining large-scale WavLM distillation with a new architecture designed for efficient streaming, ZipCodec achieves strong reconstruction and representation quality while substantially reducing the frame rate compared with existing streaming codecs. Experiments show consistent improvements over streaming baselines across reconstruction and downstream tasks, while enabling real-time inference on a consumer-grade CPU.

\section{Acknowledgments}
We gratefully acknowledge the support of NSERC, the Digital Research Alliance of Canada (alliancecan.ca), Translated
(Imminent Program), and Apple (Seed Grant) through research
funding, computing resources, and donations.

\bibliographystyle{IEEEbib}
\bibliography{refs}

\end{document}